\documentclass[preprint]{aastex}
\usepackage{graphics}

\slugcomment{To appear in the Astronomical Journal, October 2000}
\shorttitle{The AGN in IRAS~08311$-$2459}
\shortauthors{Murphy et al.}

\begin{document}

\title{The Active Nucleus in the Ultraluminous Infrared Galaxy IRAS~08311\protect\( -\protect \)2459}

\author{T. W. Murphy, Jr., B. T. Soifer\altaffilmark{1}, K. Matthews}

\affil{Palomar Observatory, California Institute of Technology, 320-47, Pasadena,
CA 91125}

\email{tmurphy@mop.caltech.edu, bts@mop.caltech.edu, kym@caltech.edu}

\and

\author{L. Armus}

\affil{SIRTF Science Center, California Institute of Technology, 314-6, Pasadena,
CA 91125}

\altaffiltext{1}{Also at the SIRTF Science Center, California Institute of Technology, 314-6, Pasadena, CA 91125}

\begin{abstract}
Near-infrared spectroscopy using the new Palomar Integral Field Spectrograph
indicates the presence of an AGN (active galactic nucleus) in the ultraluminous
infrared galaxy IRAS~08311\( - \)2459. The high-velocity wings of the Paschen-\( \alpha  \)
hydrogen recombination line are seen to be spatially unresolved, and with no
positional offset between red and blue high velocity emission. The {[}Si VI{]}
coronal line, with a 167 eV excitation potential, is also spatially unresolved
with a velocity width comparable to that of the broad component of the Pa\( \alpha  \)
emission. The low velocity component of the Pa\( \alpha  \) emission is seen
to be rotating, and is extended over \( \sim 2 \) kpc. Molecular hydrogen is
also extended, and elongated along the maximum velocity gradient, sharing the
same rotation profile as the narrow Pa\( \alpha  \) emission. The simple picture
in agreement with the observations is that of an AGN surrounded by a rotating
disk of star formation. These observations lend strength to the usage of {[}Si
VI{]} as a diagnostic of AGN activity in ultraluminous infrared galaxies, and
also highlight the utility of integral field spectroscopy in elucidating subtle
morphological differences in line emitting regions.
\end{abstract}


\newcommand{\dude}{IRAS~08311$-$2459}

\section{Introduction}

Ultraluminous infrared galaxies (ULIRGs) are among the most luminous sources
in the universe, with infrared luminosities of \( L_{ir}\ga 10^{12}L_{\odot } \).
The bulk of the energy emitted from these sources emerges in the far-infrared,
suggesting that the light we see is thermally processed by warm dust in the
galaxies. As such, the optically thick dust prohibits direct viewing of the
mechanism responsible for the extreme power generation. Quasars (or more generally
QSOs---quasi stellar objects) and massive starbursts are perhaps the only phenomena
capable of producing the high luminosities observed in these systems. Recent
spectroscopic programs at infrared wavelengths have shown that the great majority
of ULIRGs with \( L_{ir}<2\times 10^{12}L_{\odot } \) are powered predominantly
by starbursts rather than by AGN \citep{genzel,rig,twm99,vex97,vex99}. At luminosities
higher than this, the AGN fraction rises to between one-third and one-half.

\dude\ is a ULIRG at \( cz=30150 \) km~s\( ^{-1} \) (1 arcsec = 1.7 kpc) with
\( L_{ir}=2.5\times 10^{12}L_{\odot } \), and is classified as a ``warm'' ULIRG
by the definition of \citet{sand88}. This galaxy is selected from the 2Jy ULIRG
sample of \citet{str90,str92}, and is a single-nucleus galaxy surrounded by
tidal debris \citep{twm96}. The near-infrared spectrum of \dude\ \citep{twm99}
shows emission line features suggestive of AGN activity, though a scenario involving
strong shocks and high velocity outflows could also explain these spectral features.
This galaxy was selected for the present study because it is the only galaxy
in a volume-limited sample of 33 ULIRGs \citep{twm00} showing strong {[}\ion{Si}{6}{]}
emission, and one of two displaying any obvious high-velocity Pa\( \alpha  \)
emission. Calculated quantities involving luminosities and physical scales assume
a cosmology with \( H_{0}=75 \) km~s\( ^{-1} \)~Mpc\( ^{-1} \) and \( q_{0}=0 \)
throughout this paper. 

Near-infrared integral field spectroscopy, effectively providing simultaneous
velocity-resolved imaging capabilities of multiple emission lines, allows subtle
morphological differences between various line emitting regions to be elucidated.
In this paper we present integral field observations of the Pa\( \alpha  \),
H\( _{2} \)~1--0 S(3), and 167eV excitation {[}\ion{Si}{6}{]} emission lines
in \dude , which together form a consistent picture of a galaxy containing a
central AGN surrounded by a starburst disk.

\section{Observations and Data Reduction\label{obs}}

Observations were made using the Palomar Integral Field Spectrograph (PIFS)
situated at the \( f \)/70 focus of the 200-inch Hale Telescope. This instrument
produces simultaneous spectra for eight slits in a contiguous two-dimensional
field of view. A description the PIFS instrument, along with general observing
and data reduction procedures can be found in \citet{pifs}. \dude\ was observed
on the night of 25 March 1999. The \( R\approx 1300 \) resolution mode (\( \Delta v\approx 225 \)
km~s\( ^{-1} \)) was used to obtain two spectra centered on the redshifted
Pa\( \alpha  \), and H\( _{2} \)~1--0 S(3)\( + \){[}\ion{Si}{6}{]} lines,
respectively. Separate sky exposures were alternated with the on-source integrations,
with integration times of 300 s. A positional dither pattern was employed for
the sequence of integrations enabling recovery of seeing-limited spatial resolution
in the cross-slit direction. The pixel scale for these observations is 0.167
arcsec~pixel\( ^{-1} \). An offset field star was used for auto guiding, with
image motion compensated by driving the active secondary mirror. Guided observations
of a nearby star, with 10~s exposures, accompanied the spectral observations
for the purpose of evaluating the point spread function (PSF) . Wavelength calibration
is provided through a combination of OH airglow lines \citep{oo} and arc lamp
spectra taken at the time of observation. All wavelengths are referred in air
units. Atmospheric opacity and spectral flat-fielding are compensated simultaneously
using the light from HR~3862, a 4.9 mag G0~V star, spread uniformly across the
field of view. At the time of observation, the G star was at an airmass of \( \sim 1.95 \),
well matched to the airmass of the science observations. 

For these observations, the 5\farcs 4\( \times  \)9\farcs 6 field of view was
oriented with the long axis at a position angle of 90\( ^{\circ } \). The Pa\( \alpha  \)
line was observed with 1800~s of on-source integration time at an average airmass
of 2.00. The H\( _{2}+ \){[}\ion{Si}{6}{]} lines were observed for 2100~s of
on-source time at an average airmass of 1.96. The seeing for the Pa\( \alpha  \)
dataset, as determined via imaging of the PSF star, was measured at 1\farcs
00 midway through the observation, and 1\farcs 15 at the end. The telescope
was then focused, resulting in a PSF full-width at half-maximum (FWHM) of 1\farcs
05. Observations of the H\( _{2}+ \){[}\ion{Si}{6}{]} line immediately followed,
with the PSF measuring 0\farcs 85 midway through, and 0\farcs 95 at the end
of the observation. The measures of seeing should not be taken entirely at face
value, as variability over the time scales of the spectral observations limits
the accuracy of the very short PSF exposures at estimating the average seeing
conditons. Furthermore, the sliced PSF images are adequate for assessing the
approximate size of the seeing disk, but of insufficient quality to permit characterization
of subtle distortions such as may be caused by slight telescopic astigmatism.
All observations presented here were made in clear photometric conditions. Photometric
calibration was performed via \( K_{s} \) imaging of the galaxy and of the
faint standard star 9143 from \citet{persson}, using a flip-in mirror in front
of the grating, thereby establishing the continuum flux density at 2.155 \( \mu  \)m.
The continuum slopes observed in the Pa\( \alpha  \) and H\( _{2}+ \){[}\ion{Si}{6}{]}
spectra were used to estimate the continuum level for the Pa\( \alpha  \) spectrum.

Data reduction consists of first performing the sky subtraction, interpolating
static bad pixels and cosmic ray artifacts, division by the G star spectrum,
and multiplication by a blackbody spectrum matched to the G star's temperature
of 5930~K. The H\( _{2}+ \){[}\ion{Si}{6}{]} spectrum was multiplied by a template
G3~V spectrum from \citet{templates}, smoothed and resampled to the PIFS resolution,
in order to remove the Br\( \gamma  \) stellar absorption line introduced in
the division by the HR~3862 spectrum. No calibrator absorption lines of any
significance effect the Pa\( \alpha  \) spectrum. Spatial and spectral distortions
are corrected using previously generated distortion maps appropriate for the
particular grating setting. Co-registration of the eight slits in the spatial
dimension is based upon observation of the G star with its light extended perpendicular
to the slit pattern by chopping the telescope secondary mirror in a triangle-wave
pattern. The two-dimensional spectra from the eight slits are placed into the
three-dimensional datacube according to the positional dither pattern, with
a common wavelength axis established by the calibration lines. Residual OH airglow
lines are removed by subtracting a scaled version of the raw sky spectrum, with
typical scalings of \( \sim  \)2\% in absolute value, sometimes as large as
5\%. Photometric variability among individual integrations is compensated by
small scaling adjustments such that the object flux is consistent from one integration
to the next. A more complete description of the data reduction procedures may
be found in \citet{pifs}.

In the data presented here, the pure continuum images are constructed directly
from the integral field datacube, and constructed from spectral regions free
of line emission as well as OH airglow emission. For each spatial pixel, a linear
fit is made to the line-free portion of the continuum spectrum. The line images
are formed by subtracting this continuum fit from the datacube, then summing
over a small range in the spectral dimension. In this way, the line images simulate
continuum-free narrow band images at arbitrary wavelengths---or velocities---within
the datacube. The co-registration of the continuum and line images is implicit,
as the images are derived from the same dataset.

\section{Results}

\subsection{Pa\protect\( \alpha \protect \) Emission Line}

\begin{figure}[tbp]
{\par\centering \includegraphics{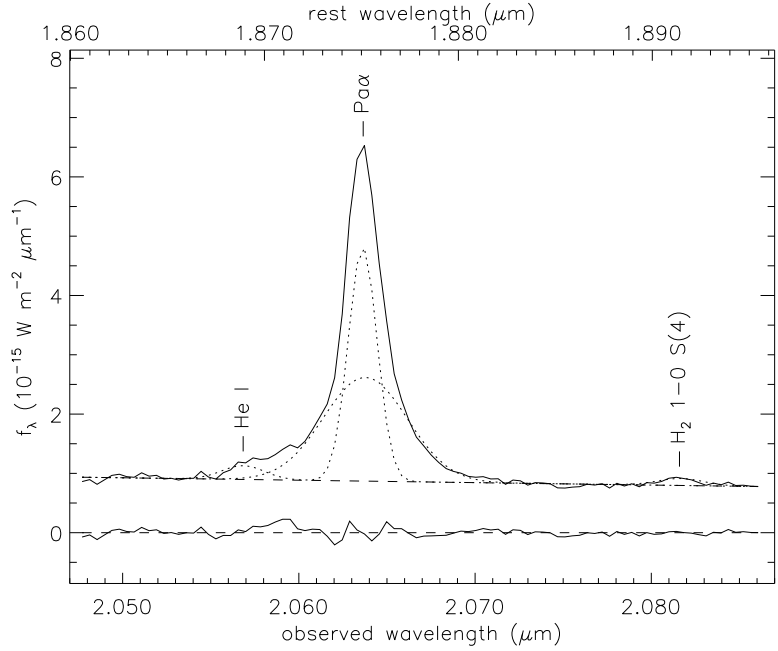} \par}

\caption{\label{fig:paspect}Nuclear spectrum of the Pa\protect\( \alpha \protect \)
line extracted from a 0\farcs 83\protect\( \times \protect \)0\farcs 83 aperture
in the datacube. Positions of the \ion{He}{1} and H\protect\( _{2}\protect \)~1--0
S(4) lines are marked at the systemic velocity determined from Pa\protect\( \alpha \protect \).
In addition to the spectral data, the continuum fit and Gaussian fit components
are shown, as is the residual to the functional fits. The broad component of
the Pa\protect\( \alpha \protect \) emission is clearly seen, and is rather
symmetric after accounting for the \ion{He}{1} line. The absolute flux density
scale corresponds to light within the aperture only. Only 18\% of the total
continuum flux from a \protect\( 5''\times 5''\protect \) box centered on the
galaxy falls within this aperture. Aperture corrections for the line emission
can be found in Table~\ref{tab:fits} of the Appendix.}
\end{figure}

Figure~\ref{fig:paspect} shows the appearance of the Pa\( \alpha  \) spectrum
(\( \lambda _{rest}=1.87510\mu  \)m), taken from a 0\farcs 83\( \times  \)0\farcs
83 square aperture centered on the continuum peak of the galaxy. The resolved
velocity profile of the line shows broad wings around a moderately narrow peak.
The wings appear asymmetric, with more emission blue-ward of the line center.
Part of this asymmetry likely comes from the \ion{He}{1} line at a rest wavelength
of 1.8689\( \mu  \)m, with an estimated contribution of \( \sim 5 \)\% of
the Pa\( \alpha  \) flux, as is characteristic in ULIRGs \citep{twm99}. Subtracting
the expected \ion{He}{1} contribution with the same velocity profile as Pa\( \alpha  \)
almost completely removes the blue excess, resulting in a Pa\( \alpha  \) profile
consistent with being symmetric. A multiple Gaussian line fit, whose components
are displayed in Figure~\ref{fig:paspect}, fits a well-centered 800 km~s\( ^{-1} \)
broad component plus a narrow, 160 km~s\( ^{-1} \) wide component to the Pa\( \alpha  \)
line, with comparable total fluxes. See Appendix~\ref{append} and Table~\ref{tab:fits}
for details on the line fits. A small amount of excess emission appears in the
fit residuals of Figure~\ref{fig:paspect} on the blue side of the Pa\( \alpha  \)
line. This emission only amounts to \( \sim  \)4\% of the flux of the broad
Gaussian component, which was not substantial enough to encourage a broad component
fit that had any asymmetric quality (i.e., blue-shifted line center). Therefore,
while this emission is real, it is of little importance in comparison to the
bulk of the broad emission. Throughout this paper we refer to this 800 km~s\( ^{-1} \)
emission feature as the ``broad'' component, though it does not represent a
canonical broad line region. We simply use this term to indicate velocities
larger than would be expected in ordinary galaxy kinematics, such as rotation.

\begin{figure}[tbp]
{\par\centering \includegraphics{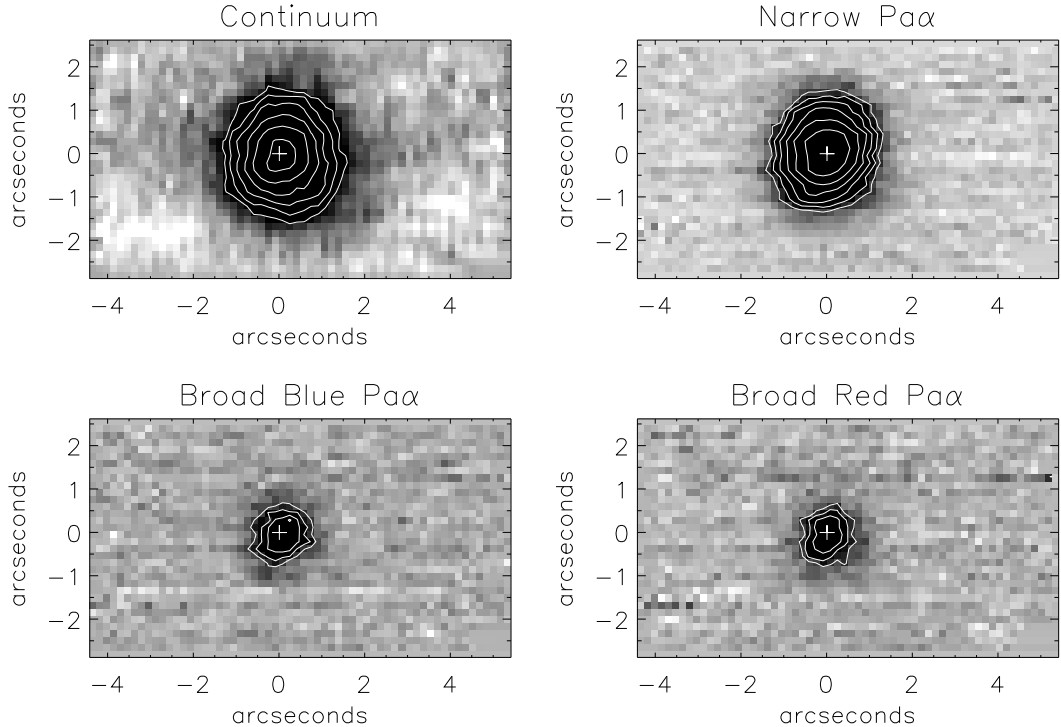} \par}

\caption{\label{fig:papicts}Morphologies of the continuum and of the Pa\protect\( \alpha \protect \)
emission in three velocity bins, centered at \protect\( v=0\protect \) and
\protect\( v=\pm 425\protect \) km~s\protect\( ^{-1}\protect \), each 250
km~s\protect\( ^{-1}\protect \) wide. All show rather round, symmetric morphologies,
though the broad emission is significantly more compact than either the narrow
Pa\protect\( \alpha \protect \) or continuum emission. The FWHM of the broad
line emission measures \protect\( \sim \protect \)1\farcs 1, compared to 1\farcs
55 and 1\farcs 75 for the narrow Pa\protect\( \alpha \protect \) and continuum
emission, respectively. North is up, and east is to the left in all images.
Contours are placed at \protect\( \sqrt{2}\protect \) multiplicative intervals.}
\end{figure}

The continuum morphology is presented in Figure~\ref{fig:papicts}, along with
images of three different spectral ranges within the Pa\( \alpha  \) line profile.
Referring to the line peak at 2.06345\( \mu  \)m as the zero velocity reference,
the three continuum-subtracted narrow band images correspond to central velocities
of \( -425 \), 0, and \( +425 \) km~s\( ^{-1} \), each 250 km~s\( ^{-1} \)
wide. The narrow line image is more compact than the continuum, with characteristic
FWHM values of 1\farcs 55 and 1\farcs 75, respectively. The line emission is
also slightly elongated compared to the round continuum, with an ellipticity
(major-to-minor axis ratio) of \( \sim 1.1 \). The two images of the line wings
are significantly more compact than either the continuum or narrow line Pa\( \alpha  \)
images, and are in fact consistent with the estimated 1\farcs 1 seeing limit
at the time of observation. No significant difference in shape, size, or center
can be distinguished between the red and blue line wing images, though they
each exhibit an ellipticity of \( \sim 1.12 \). The observed ellipticity could
be the result of a slightly defocused, astigmatic telescope. Such a distortion
would not noticeably distort the much larger continuum shape, and likewise would
not fully account for the similar elongation of the narrow Pa\( \alpha  \)
image.

\begin{figure}[tbp]
{\par\centering \includegraphics{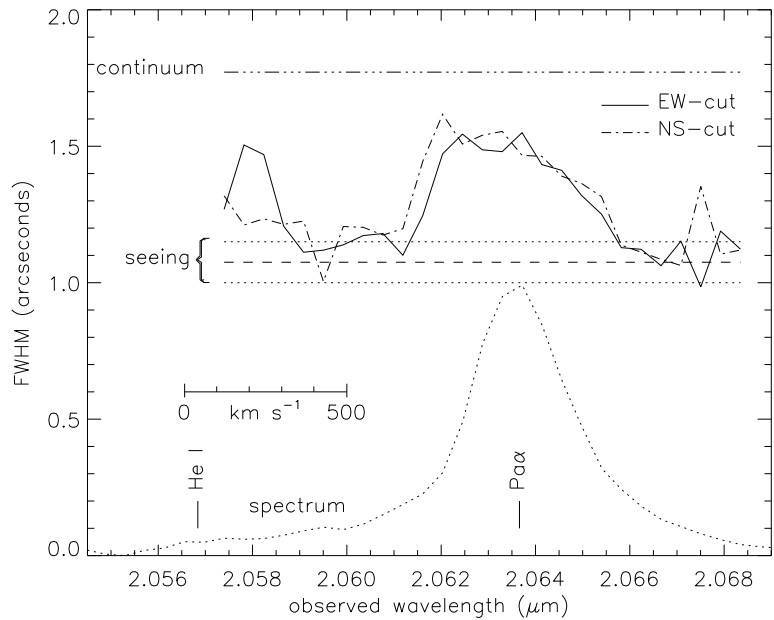} \par}

\caption{\label{fig:pafwhm}Spatial line extent as a function of wavelength for the
Pa\protect\( \alpha \protect \) line, showing the spatially resolved nature
of the narrow line emission in contrast to the spatially unresolved high velocity
regions. Two plots are shown, representing measures in the orthogonal array
directions. The size of the continuum source is shown at the top of the plot.
The estimated range in seeing is indicated, as measured from PSF images intermixed
with the spectral data. The profile at bottom shows the continuum subtracted
spectrum of the Pa\protect\( \alpha \protect \) line from Figure~\ref{fig:paspect}
for immediate comparison to the spatial extents. The increase in spatial width
at the left edge of the plot is attributable to the \ion{He}{1} line. Line centers
are marked for the systemic velocity, and a velocity scale is provided for convenient
reference.}
\end{figure}

With confidence in the generic morphological character of the line emission,
we may turn to a more detailed representation of the line extents by plotting
the line emission FWHM as a function of wavelength. Figure~\ref{fig:pafwhm}
shows the measured spatial FWHM in both the \( x \) and \( y \) dimensions
of the array, corresponding to the east--west and north--south directions. Each
data point represents a measurement in the spatial plane at the location of
a single spectral pixel. Because a spectral resolution element spans roughly
four pixels, adjacent points in Figure~\ref{fig:pafwhm} are not independent.
The horizontal dotted lines delimit the range of atmospheric seeing measured
in conjunction with the spectral observations, with a dashed line indicating
the mean of these two values. The extended nature of the narrow line emission
stands out in clear contrast to the seeing-limited extent of the line wings.
The increase in width at the left edge of the plot coincides with the red side
of the \ion{He}{1} emission line. Due to the low signal-to-noise ratio in the
region of the \ion{He}{1} line, it was not possible to extend the plot data
to shorter wavelengths. The continuum FWHM, measured beyond the line wings,
is represented by the dashed line at top, and the continuum-subtracted line
profile is plotted for convenient reference at bottom.

\subsection{H\protect\( _{2}\protect \)~1--0 S(3) and {[}\ion{Si}{6}{]} Emission Lines}

\begin{figure}[tbp]
{\par\centering \includegraphics{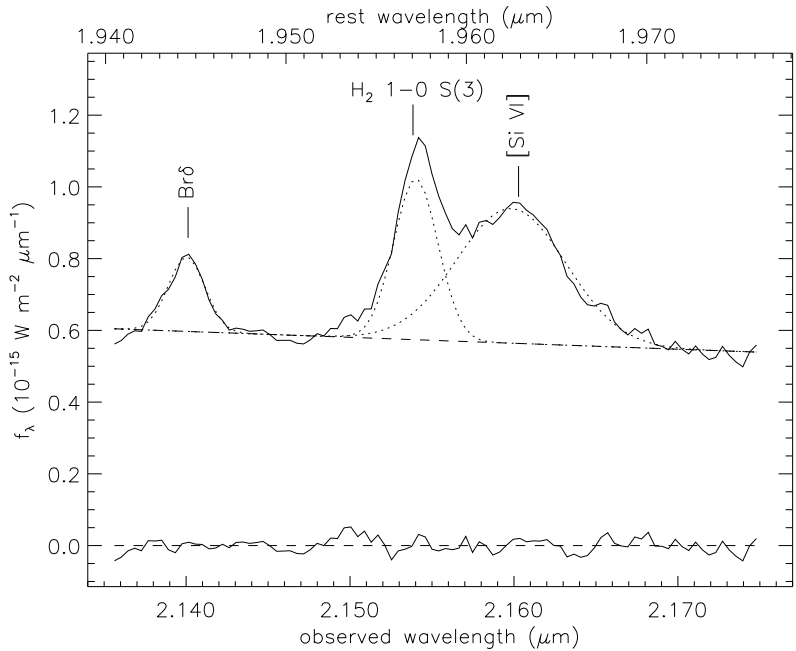} \par}

\caption{\label{fig:h2spect}Nuclear spectrum of the H\protect\( _{2}\protect \)~1--0
S(3) \protect\( +\protect \) {[}\ion{Si}{6}{]} line region, extracted from
a 0\farcs 67\protect\( \times \protect \)0\farcs 67 aperture in the datacube.
The Br\protect\( \delta \protect \) line is also seen. In addition to the spectral
data, the continuum fit and Gaussian fit components are shown, as is the residual
to the functional fit. The {[}\ion{Si}{6}{]} line is clearly broader than the
other lines in the spectrum, and comparable in strength to the H\protect\( _{2}\protect \)
line. Line centers are marked at the systemic velocity. The absolute flux density
scale corresponds to light within the aperture only. Only 13.5\% of the total
continuum flux from a \protect\( 5''\times 5''\protect \) box centered on the
galaxy falls within this aperture. Aperture corrections for the various emission
lines can be found in Table~\ref{tab:fits} of the Appendix.}
\end{figure}

The integral field spectrum of \dude\ at 2.15\( \mu  \)m contains the Br\( \delta  \),
H\( _{2} \)~1--0 S(3), and {[}\ion{Si}{6}{]} emission lines at rest wavelengths
of 1.94456\( \mu  \)m, 1.95702\( \mu  \)m, and 1.96287\( \mu  \)m, respectively.
Figure~\ref{fig:h2spect} shows the one-dimensional spectrum, extracted from
a 0\farcs 67\( \times  \)0\farcs 67 square aperture centered on the continuum
peak. The Br\( \delta  \) line is largely ignored here, as the Pa\( \alpha  \)
line is a far more effective probe of hydrogen recombination, and the combination
of Br\( \delta  \)'s weakness plus its spectral proximity to Pa\( \alpha  \)
limits its usefulness as a measure of extinction. The 1\farcs 35 spatial extent
of the Br\( \delta  \) line is perfectly consistent with the Pa\( \alpha  \)
extent, accounting for the seeing difference between the observations. The {[}\ion{Fe}{2}{]}
line at 1.9670 \( \mu  \)m, seen in some ULIRGs \citep{twm99}, does not appear
in the spectrum, though the Br\( \gamma  \) feature from the atmospheric calibrator
coincides with this same wavelength, making the presence of {[}\ion{Fe}{2}{]}
uncertain.

All lines in the H\( _{2}+ \){[}\ion{Si}{6}{]} spectrum are spectrally resolved,
with the H\( _{2} \) line resembling the atomic hydrogen lines in terms of
observed line width. Though the H\( _{2} \) line is blended with the {[}\ion{Si}{6}{]}
line, it is known to be symmetric by comparing with the H\( _{2} \)~1--0 S(1)
line profile \citep{twm99}. The {[}\ion{Si}{6}{]} line is significantly broader
than its neighboring lines, with a FWHM of \( \sim 1000 \) km~s\( ^{-1} \).
See Appendix~\ref{append} and Table~\ref{tab:fits} for more information on
the functional fits to these lines.

\begin{figure}[tbp]
{\par\centering \includegraphics{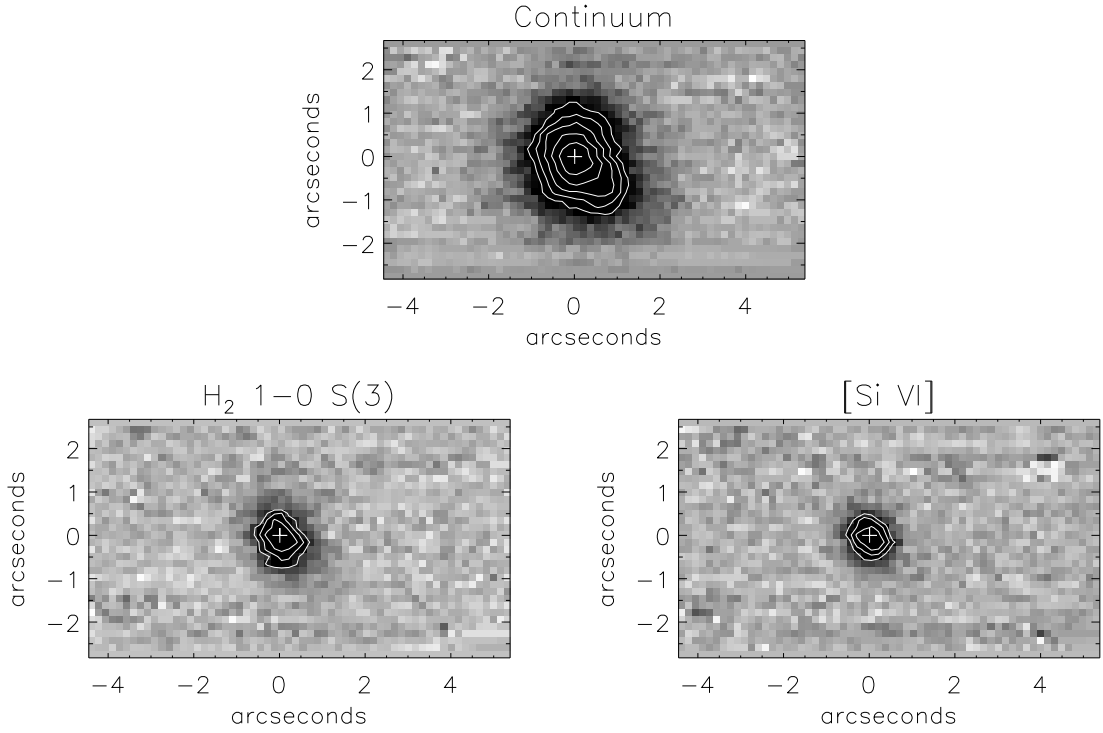} \par}

\caption{\label{fig:h2picts}Continuum and line morphologies for the H\protect\( _{2}\protect \)
and {[}\ion{Si}{6}{]} lines. The line images are each 400 km~s\protect\( ^{-1}\protect \)
wide, centered on the line peaks. The {[}\ion{Si}{6}{]} emission is completely
unresolved, as well as symmetric. The H\protect\( _{2}\protect \) line is slightly
resolved, and extended along a 30\protect\( ^{\circ }\protect \) position angle,
roughly consistent with the disk direction inferred from the rotation. The continuum
shape differs slightly from that seen in Figure~\ref{fig:papicts}, in that
a small extension appears to the SW. The seeing conditions were better for this
image, and the compact symmetry of the {[}\ion{Si}{6}{]} line suggests that
this extension is real. Orientation and contours are as described for Figure~\ref{fig:papicts}.}
\end{figure}

Figure~\ref{fig:h2picts} displays the continuum and narrow band line images
of \dude\ in H\( _{2} \)~1--0 S(3) and {[}\ion{Si}{6}{]} emission. The continuum
morphology appears different in Figures~\ref{fig:papicts} and \ref{fig:h2picts},
mainly due to 20\% better seeing and telescope focus in the latter (see Section~\ref{obs}).
In Figure~\ref{fig:h2picts} the continuum appears to have a weak extension
to the SW. The line images are clearly more compact than the continuum, and
the near perfect symmetry of the {[}\ion{Si}{6}{]} emission suggests that the
structure seen in the continuum is real, and simply not resolved in Figure~\ref{fig:papicts}.
Indeed the central contours of the continuum image in Figure~\ref{fig:h2picts}
appear much more symmetric than the outer contours. The line images are created
in the same manner as those in Figure~\ref{fig:papicts}, with the spectral
range encompassing points within \( \pm 200 \) km~s\( ^{-1} \) of the line
center. The {[}\ion{Si}{6}{]} emission line is slightly more compact than the
neighboring H\( _{2} \) line. An elliptical Gaussian fit to each of these line
images yields FWHM sizes (average of major and minor axes) of 1\farcs 1 for
H\( _{2} \) and 0\farcs 9 for {[}\ion{Si}{6}{]}, with ellipticities (major-to-minor
axis ratios) of 1.30 and 1.03, respectively. The H\( _{2} \) elongation appears
to be real, with a position angle on the sky of \( \sim 30^{\circ } \)---similar
to the disk orientation implied by the rotation axis, discussed in Section~\ref{rotation}.

\begin{figure}[tbp]
{\par\centering \includegraphics{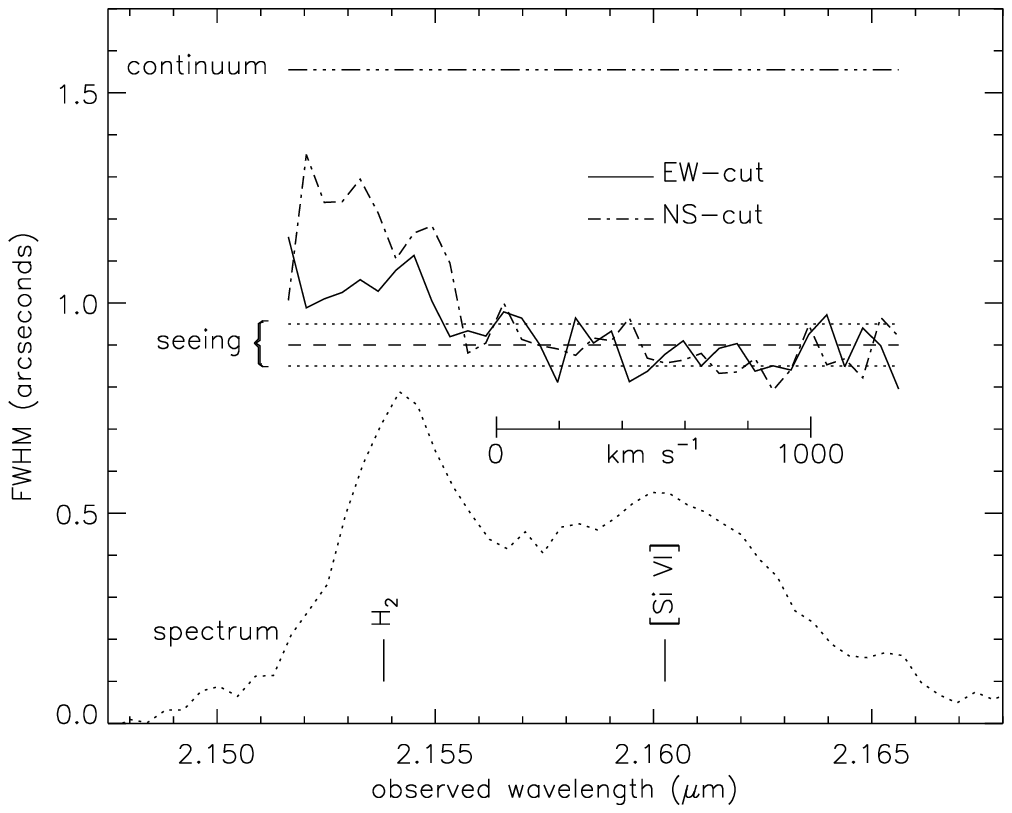} \par}

\caption{\label{fig:h2fwhm}Spatial line extent as a function of wavelength for the
H\protect\( _{2}\protect \)~1--0 S(3) and {[}\ion{Si}{6}{]} lines, showing
the spatially unresolved nature of the {[}\ion{Si}{6}{]} emission in contrast
to the spatially resolved H\protect\( _{2}\protect \) emission. See the caption
for Figure~\ref{fig:pafwhm} for a description of plot features. The spectrum
at bottom is a continuum subtracted version of that seen in Figure~\ref{fig:h2spect}.}
\end{figure}

A more detailed look at the difference in spatial distribution between H\( _{2} \)
and {[}\ion{Si}{6}{]} can be seen in Figure~\ref{fig:h2fwhm}, which, like Figure~\ref{fig:pafwhm},
plots the spatial FWHM of the line distribution in the east--west and north--south
directions as a function of wavelength. Referring to the line profile at the
bottom of Figure~\ref{fig:h2fwhm}, it is clear that throughout the wavelength
range corresponding to the broad {[}\ion{Si}{6}{]} emission, both the \( x \)
and \( y \) widths are consistent with the seeing limit, indicating that the
entire {[}\ion{Si}{6}{]} emission line is spatially unresolved. The similarity
in widths in the two orthogonal directions further demonstrates the symmetry
of the PSF during the observation. On the short wavelength side of the {[}\ion{Si}{6}{]}
emission, the spatial profile broadens, marking the location of the H\( _{2} \)
line. Here, the north--south extent is larger than the east--west extent, owing
to the elongated nature of the emission at a 30\( ^{\circ } \) position angle.

\subsection{Rotation of H\protect\( _{2}\protect \) \& Narrow Pa\protect\( \alpha \protect \)\label{rotation}}

\begin{figure}[tbp]
{\par\centering \includegraphics{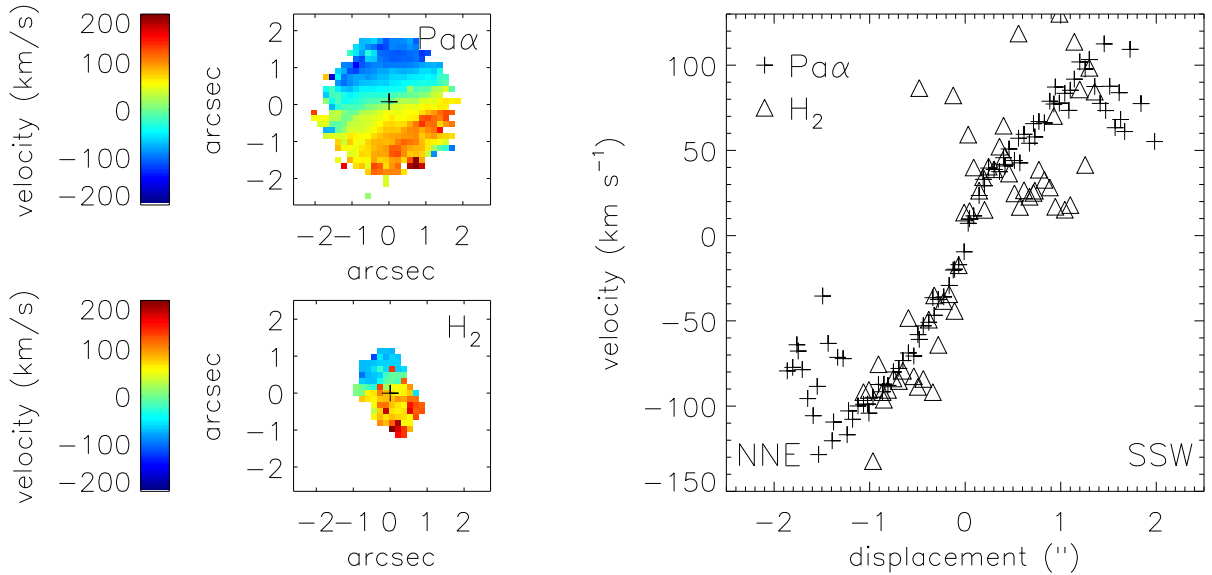} \par}

\caption{\label{fig:rot}Velocity fields of both the Pa\protect\( \alpha \protect \)
and H\protect\( _{2}\protect \) lines. To the left are color images representing
the velocity of the peak emission, with red indicating redshift, etc. A clear
rotational signature is seen in both. To the right is a position-velocity plot
for both species along a 0\farcs 67 ``slit'' at a position angle of 20\protect\( ^{\circ }\protect \),
corresponding to the angle of maximum velocity gradient. The Pa\protect\( \alpha \protect \)
points (plus symbols) trace out a very clean rotation pattern, and the noisier
H\protect\( _{2}\protect \) points (triangles) follow the same profile.}
\end{figure}

Taking advantage of the coexistence of two-dimensional spatial plus spectral
information, we can construct velocity fields of the line emitting gas. At each
spatial pixel, the wavelength of the peak line emission is computed, and converted
to a velocity. Figure~\ref{fig:rot} displays the velocity fields obtained in
this manner for \dude\ in both the Pa\( \alpha  \) and H\( _{2} \)~1--0 S(3)
lines. Both lines exhibit a velocity structure indicative of pure rotation.
The color scale is the same for both maps, through which it is seen that the
rotation is not only aligned along the same axis for the two lines, but with
roughly the same amplitude. The rotation axis is at a position angle of \( \sim 110^{\circ } \),
implying a projected disk major axis of \( \sim 20^{\circ } \), consistent
with the observed elongation observed for H\( _{2} \). Figure~\ref{fig:rot}
also shows the rotation curves obtained for both lines, in a 0\farcs 67 wide
aperture through the galaxy center at a 20\( ^{\circ } \) position angle. Though
the molecular hydrogen data has much more scatter than does the Pa\( \alpha  \)
curve, it is clear that the two constituents---the narrow Pa\( \alpha  \) and
the H\( _{2} \) emission---share the same physical rotation.

\section{Discussion}

Combining visible light spectroscopic surveys of ULIRGs \citep[e.g.,][]{kvs98,vks99}
with infrared observations \citep{vex97,vex99}, it is found that 20--25\% of
ULIRGs can be characterized as containing AGN. The visible studies generally
utilize the {[}\ion{O}{3}{]}/H\( \beta  \) line ratio as the primary diagnostic
in differentiating Seyfert (AGN) galaxies from starburst-dominated galaxies,
while infrared identification of AGN are generally based on the presence of
broad atomic recombination lines, and occasionally of {[}\ion{Si}{6}{]}. ULIRGs
are known to sometimes harbor galactic-scale superwinds \citep{ham}, which could
potentially contribute to the observed broad line wings and to {[}\ion{Si}{6}{]}
emission through powerful shock excitation. The present integral field data
for \dude\ argue for a simple picture of a compact AGN surrounded by a more
diffuse starburst---at least some of which is occurring in a rotating disk around
the nucleus. While massive outflows and expanding shells are present in some
ULIRGs, they may not be ubiquitous. The PIFS data on \dude\ highlights the important
role that integral field spectroscopy can play in untangling the dynamics and
excitation of the circumnuclear gas in ULIRGs.

Without considering the contribution of the \ion{He}{1} line to the Pa\( \alpha  \)
line profile, one might conclude that the apparently blue-asymmetric profile
of Pa\( \alpha  \) indicates the presence of outflow phenomena, as is often
the case for asymmetric line profiles \citep[e.g.,][]{ham,wang}. Knowledge of
the \ion{He}{1} line on the blue side of Pa\( \alpha  \), with expectations
on typical ULIRG line strengths \citep{twm99,twm00} allow us to understand the
Pa\( \alpha  \) profile as the combination of a symmetric Pa\( \alpha  \)
line mixed with a similarly shaped \ion{He}{1} line. The Gaussian line fits
give the \ion{He}{1} line 4\% of the total Pa\( \alpha  \) flux---consistent
with expectations from the ULIRG population (\citeauthor{twm99}). Perhaps the
strongest evidence that the \ion{He}{1} line is contributing to the total line
profile comes from the observation of increased spatial extent coincident with
the red side of the \ion{He}{1} line, shown in Figure~\ref{fig:pafwhm}.

While we can not rule out the presence of a very compact nuclear outflow, the
symmetry of the Pa\( \alpha  \) line profile resulting from proper accounting
of the \ion{He}{1} flux, together with the fact that both sides of the profile's
broad wings are spatially unresolved (\( <500 \) pc) in the full two-dimensional
sense, argue that the broad emission does not arise from an outflow, or wind,
phenomenon. Additionally, the red and blue wing components show no relative
displacement greater than the 0\farcs 05 level, or \( \sim 100 \) pc, further
supporting this conclusion. If a wind is to account for the high-velocity Pa\( \alpha  \)
emission, the lack of significant extinction to the red emission (\( A_{V}<0.5 \)
mag based on line symmetry) must be explained, as must the fact that the physical
scale is smaller than the size of the narrow line region, which is spatially
resolved. We believe the broad emission arises from the near vicinity of an
obscured AGN, and that this component is separate from the rotating, narrow
emission discussed at the end of this section. 

The presence of {[}\ion{Si}{6}{]} in the spectrum of \dude\ is a clear indicator
of high energy processes at work in the galaxy. {[}\ion{Si}{6}{]} is not commonly
seen in ULIRGs, even those exhibiting broad Pa\( \alpha  \) emission \citep{twm99}.
\citet{vex97,vex99} see evidence for {[}\ion{Si}{6}{]} in a few ULIRGs, though
blending with the comparably strong H\( _{2} \)~1--0 S(3) line prohibits accurate
estimates of flux contributions. The {[}\ion{Si}{6}{]} emission in \dude\ measures
\( 6\times 10^{7}L_{\odot } \) in luminosity, uncorrected for extinction. This
is among the most luminous {[}\ion{Si}{6}{]} emission line yet observed, although
relative to the bolometric luminosity only slightly more luminous than that
found in nearby Seyfert galaxies \citep{marco}. The {[}\ion{Si}{6}{]}-to-bolometric
luminosity ratio for quasars is the same as that seen in \dude , based on the
template QSO spectrum in \citeauthor{twm99}

Main sequence stars are incapable of producing significant flux at 167 eV---the
requirement for creating Si\( ^{5+} \). Only ionizing radiation from an AGN
or fast shocks can account for the {[}\ion{Si}{6}{]} emission \citep{contini}.
It is difficult to differentiate between these excitation processes observationally,
though a diagnostic utilizing the {[}\ion{Fe}{7}{]} \( \lambda 0.6087\mu  \)m
line was explored by \citet{marco}, in which all instances of {[}\ion{Si}{6}{]}
detections in Seyfert galaxies are attributed to photoionization.

The integral field data may contribute a clue to the nature of the {[}\ion{Si}{6}{]}
excitation. The molecular hydrogen emission---often associated with collisional
excitation---has a significantly different morphological character than the
{[}\ion{Si}{6}{]} emission. Namely, the H\( _{2} \) line originates from a
region roughly one kpc larger than that of {[}\ion{Si}{6}{]}. The underlying
morphology of violent shocks in major mergers is not clearly understood. However,
if the {[}\ion{Si}{6}{]} line was shock excited---perhaps tracing out a physical
boundary along a shock front---then any molecular hydrogen excited by this same
shock phenomenon would arise from the near vicinity of the shock front \citep{shocks},
thereby sharing similar spatial structure with the {[}\ion{Si}{6}{]} emission.
Naturally, the very strong shocks capable of producing Si\( ^{+5} \) would
probably dissociate H\( _{2} \) molecules, but the precursor and post-shock
mechanisms may be capable of producing significant H\( _{2} \) emission. In
order for both the {[}\ion{Si}{6}{]} and H\( _{2} \) emission to be shock-excited
requires two separate shock components with vastly different velocities and
spatial scales. While we can in no way rule out this possibility with the current
data, a simpler explanation has the {[}\ion{Si}{6}{]} emission originating from
photoionization by a central AGN, with the H\( _{2} \) emission excited by
other means.

Further differentiating the H\( _{2} \) distribution from that of {[}\ion{Si}{6}{]}
is the fact that the H\( _{2} \) emission is extended along the direction of
maximum velocity gradient---the inferred disk axis. A simple quadrature decomposition
of the H\( _{2} \) 1\farcs \( 0\times 1 \)\farcs 3 FWHM ellipse with a 0\farcs
9 seeing disk yields an intrinsic aspect ratio of greater than 2:1, with a long
axis extent of about 1.6 kpc. The rotation curve derived from the H\( _{2} \)
emission is in complete agreement with that seen for Pa\( \alpha  \) (Figure~\ref{fig:rot}),
suggesting that the narrow Pa\( \alpha  \) and H\( _{2} \) emission arise
from the same rotating disk. Similar observations of elongated molecular gas
concentrations coincident with disk orientation and rotation are also seen in
CO data of luminous infrared galaxies \citep{bryant}. 

A picture emerges wherein both the \ion{H}{1} and H\( _{2} \) narrow line emission
arises from a rotating, star forming disk. Alternatives for production of the
narrow line emission include X-ray photoionization of the disk from the central
source, or photoionization of the gas from ``precursor'' radiation emanating
from a powerful shock front. Both of these alternatives suffer from the same
difficulty of requiring the ionizing radiation to penetrate almost 1 kpc through
the disk in order to produce the large scale emission observed. If indeed star
formation is occurring on a scale of \( \sim 1.5 \) kpc---as the H\( _{2} \)
emission indicates---then \dude\ boasts star formation on a significantly larger
scale than the 100--300 pc typically seen among ULIRGs in mid-infrared studies
\citep{soifer}. 

If the narrow, rotating component of the Pa\( \alpha  \) line is produced by
young stars in a disk, the implied star formation rate (SFR) is \( \sim 75 \)
\( M_{\odot } \)~yr\( ^{-1} \), assuming an intrinsic H\( \alpha / \)Pa\( \alpha  \)
line flux ratio of 8.6 \citep[][Case B with $n_e=10^4$ cm$^{-3}$ and $T=10000$ K]{ost}
and a conversion between H\( \alpha  \) luminosity and SFR following \citet{kennicutt}.
This estimate only includes the flux in the narrow Gaussian component to the
Pa\( \alpha  \) line, as portrayed in Figure~\ref{fig:paspect}. Of course,
this conversion takes into account neither the extinction at Pa\( \alpha  \),
nor the contribution of the central AGN to the narrow Pa\( \alpha  \) line
flux---both of which may be significant in this system, yet would affect our
SFR estimate in opposite senses. Unpublished visible light spectra of \dude\
indicate an extinction to the disk of \( A_{V}\approx 1.1 \) mag, comparing
the Pa\( \alpha  \) and H\( \beta  \) emission, leading to 0.16 mag of extinction
at Pa\( \alpha  \). The AGN contribution may exceed this value, such that the
above SFR estimate likely represents an upper limit.

If the entire far-infrared luminosity in \dude\ were attributable to star formation,
then one may compute the expected total star formation rate by two methods.
The first, from \citet{sco}, puts a lower bound on the bolometric luminosity
by assuming that the luminosity is dominated by stars converting protons to
He via the CNO cycle, and that the rate of mass reduction is the rate at which
stars are forming. This approach yields an expected SFR of \( \sim 200 \) \( M_{\odot } \)~yr\( ^{-1} \).
The other method, from \citet{hunter}, integrates the total stellar luminosity
output to produce an estimated SFR of 650 \( M_{\odot } \)~yr\( ^{-1} \),
assuming all of the stellar luminosity is reprocessed into infrared emission
(\( \beta =1 \) in their model). Therefore, it is likely that either there
is a large amount of star formation that remains obscured at 2 \( \mu  \)m,
or that the hidden AGN is a substantial contributor to the infrared luminosity
in this system. Given the presence of the unresolved, relatively strong {[}\ion{Si}{6}{]}
emission we favor the latter interpretation, though with an estimated SFR of
\( \sim  \)75 \( M_{\odot } \) yr\( ^{-1} \), star formation could clearly
contribute a significant portion of the total luminosity.

\section{Conclusions}

Integral field spectroscopy of the suspected AGN-powered ULIRG, \dude , provides
important constraints on the nature of the line emitting regions in this galaxy.
The evidence presented can be reasonably explained by the presence of a bona-fide
quasar embedded within a disk of star formation in the post-merger system. Both
components contribute significantly to the total energy production in the galaxy.
The salient points derived from these data are:

\begin{enumerate}
\item The low velocity emission of Pa\( \alpha  \), along with H\( _{2} \), has
been identified with a rotating disk of material with a diameter of \( \sim 1.5 \)
kpc. It is likely that widespread star formation is responsible for the observed
narrow line emission from this region.
\item Moderately broad emission is seen in both Pa\( \alpha  \) and {[}\ion{Si}{6}{]},
with a projected FWHM\( \approx 1000 \) km~s\( ^{-1} \). The broad emission
is spatially unresolved (\( <500 \) pc) in both species. Both lines are spectrally
symmetric, and there is no observed spatial offset between the red and blue
wings of the broad Pa\( \alpha  \) emission. Outflow phenomena are therefore
unlikely to account for the high velocity emission, with an AGN being the preferred
source of the {[}\ion{Si}{6}{]} and broad Pa\( \alpha  \) emission.
\end{enumerate}

\acknowledgements

We thank Michael Strauss for his role in the early stages of the Caltech effort
in studying ULIRGs. The wavelength reference for the {[}\ion{Fe}{2}{]} line
was provided by James Graham. We also thank the night assistant at Palomar,
Rick Burruss, for assistance in the observations. T.W.M. is supported by the
NASA Graduate Student Researchers Program, and the Lewis Kingsley Foundation.
This research is supported by a grant from the National Science Foundation.

\appendix

\section{Functional Line Fits\label{append}}

Functional line fits have been applied to the extracted nuclear spectra of \dude
, and displayed in Figures~\ref{fig:paspect} and \ref{fig:h2spect}. In each
case, the continuum is fit with a linear function, and all lines except for
Pa\( \alpha  \) are fit with a single Gaussian profile. Attempting to fit Pa\( \alpha  \)
with a single Gaussian resulted in an obviously inadequate fit, but the resulting
parameters are useful in fitting weaker atomic lines. The signal-to-noise ratio
in the weaker lines does not justify the use of multiple component fits, yet
their underlying profiles are expected to closely resemble Pa\( \alpha  \).
The single-fit Pa\( \alpha  \) has a central velocity of 30147 km~s\( ^{-1} \)
and a FWHM of 415 km~s\( ^{-1} \); accounting for an instrumental resolution
of 230 km~s\( ^{-1} \), this translates to an intrinsic line width of 345 km~s\( ^{-1} \).
These numbers become the constraints used, where necessary, for the low ionization
atomic lines.

The Pa\( \alpha  \) spectrum contains three observable emission features, fit
with four Gaussian profiles---two for Pa\( \alpha  \) itself. It was found
to be necessary to fix the \ion{He}{1} line center and width according to the
single-Gaussian Pa\( \alpha  \) values as discussed above, leaving only the
flux as a free parameter. The two components in the Pa\( \alpha  \) line fit
were unconstrained, and tended to repeatably settle to the same parameters despite
factors of two modifications to the initial estimated values. The H\( _{2} \)~1--0
S(4) line was constrained to share the 380 km~s\( ^{-1} \) line width found
for the H\( _{2} \)~1--0 S(3) line, as discussed below. The plot of fit residuals
at the bottom of Figure~\ref{fig:paspect} shows that the two-Gaussian fit for
Pa\( \alpha  \) does a passable job, but excess emission on the blue side well
above the noise level demonstrates the approximate nature of the fitting business.
For this reason, the separate components in the Pa\( \alpha  \) fit are not
to be taken as individual, distinct physical components, but rather as a tool
for gaining a qualitative feel for the nature and symmetry of the broad emission.
Table~\ref{tab:fits} lists the line properties and their fit parameters.

\begin{deluxetable}{lccccc}
\tabletypesize{\footnotesize}
\tablewidth{0pt}
\tablenum{1} 
\tablecaption{Fit Parameters and Line Fluxes\label{tab:fits}}
\tablehead{
\colhead{Line} & \colhead{$cz-30150$} & 

\colhead{FWHM\tablenotemark{a}} & \colhead{Equiv. Width} & 

\colhead{Aperture\tablenotemark{b}} & 

\colhead{Total Flux\tablenotemark{c}} \\
& \colhead{(km~s$^{-1}$)} & \colhead{(km~s$^{-1}$)} & 

\colhead{(nm)} & \colhead{Correction} & 

\colhead{($\times 10^{-18}$ W~m$^{-2}$)}
} 
\startdata
\ion{He}{1} & \phs\phn 0\tablenotemark{d} & \phn 339\tablenotemark{d} &

\phn 0.78 & 0.21\tablenotemark{e} & \phn 3.6 \\ 
Pa$\alpha$\hfill narrow & \phn$-9$ & \phn 156 & \phn 9.3\phn & 

0.19\tablenotemark{f} & 45.9 \\
\hfill broad & \phs 16 & \phn 779 & 12.0\phn & 0.30\tablenotemark{f} 

& 34.1 \\
H$_2$ 1--0 S(4) & \phs\phn 7 & \phn 376\tablenotemark{d} 

& \phn 0.48 & 0.26\tablenotemark{e} & \phn 1.7 \\
Br$\delta$ & $-14$ & \phn 297 & \phn 0.97 & 0.15\tablenotemark{f} & 

\phn 3.5 \\
H$_2$ 1--0 S(3) & \phs 28 & \phn 384 & \phn 2.6\phn & 

0.21\tablenotemark{f} & \phn 7.4 \\
$[$\ion{Si}{6}$]$ & $-66$ & 1066 & \phn 5.6\phn & 0.30\tablenotemark{f} 

& 10.8 \\
\enddata 
\tablenotetext{a}{Deconvolved via simple quadrature by the instrument resolution of 230 km~s$^{-1}$ in the Pa$\alpha$ spectrum, and 220 km~s$^{-1}$ in the H$_2 +[$\ion{Si}{6}$]$ spectrum.}
\tablenotetext{b}{Ratio of flux within extraction aperture (i.e., spectra presented in Figures~\ref{fig:paspect} and \ref{fig:h2spect}) to total line flux encompassing detected extent of line emission. For strong lines with spectrally distinct spatial characteristics (as determined from Figures~\ref{fig:pafwhm} and \ref{fig:h2fwhm}), this ratio is measured directly from the datacube.}
 \tablenotetext{c}{Aperture-corrected flux scaled from line fit, within 5 arcsec aperture; 15\% estimated error.}
\tablenotetext{d}{Held fixed in line fit.}
\tablenotetext{e}{Calculated based on line FWHM and aperture size assuming symmetric Gaussian spatial profile.}
\tablenotetext{f}{Measured directly from datacube.}
\end{deluxetable}

The H\( _{2}+ \){[}\ion{Si}{6}{]} spectrum contains three lines, each fit with
a single Gaussian profile. The fit was performed placing no constraints on the
three parameters of each line fit. Figure~\ref{fig:h2spect} shows the result
of the combined fit, and Table~\ref{tab:fits} gives the properties of the individual
line fits.

\clearpage

\end{document}